\documentclass[conference]{IEEEtran}
\IEEEoverridecommandlockouts
\usepackage{amsfonts} % added
\usepackage{amsmath} % added
\newcommand{\quotes}[1]{``#1''} %defined
\usepackage{multirow}

\usepackage{booktabs}
\usepackage{subfigure}
\usepackage{url}

\usepackage{cite}
\usepackage{amsmath,amssymb,amsfonts}
\usepackage{algorithmic}
\usepackage{graphicx}
\usepackage{textcomp}
\usepackage{xcolor}
\def\BibTeX{{\rm B\kern-.05em{\sc i\kern-.025em b}\kern-.08em
    T\kern-.1667em\lower.7ex\hbox{E}\kern-.125emX}}
\begin{document}

\title{Pre-trained Transformer Uncovers Meaningful Patterns in Human Mobility Data}

\author{\IEEEauthorblockN{Alameen Najjar}
\IEEEauthorblockA{\textit{Rakuten Institute of Technology} \\
Tokyo, Japan \\
alameen.najjar@rakuten.com}
}

\IEEEoverridecommandlockouts
\IEEEpubid{\makebox[\columnwidth]{ \hfill}
% \IEEEpubid{\makebox[\columnwidth]{}
\hspace{\columnsep}\makebox[\columnwidth]{ }}
\maketitle
\IEEEpubidadjcol

\begin{abstract}
We empirically demonstrate that a transformer pre-trained on country-scale unlabeled human mobility data learns embeddings capable, through fine-tuning, of developing a deep understanding of the target geography and its corresponding mobility patterns. Utilizing an adaptation framework, we evaluate the performance of our pre-trained embeddings in encapsulating a broad spectrum of concepts directly and indirectly related to human mobility. This includes basic notions, such as geographic location and distance, and extends to more complex constructs, such as administrative divisions and land cover. Our extensive empirical analysis reveals a substantial performance boost gained from pre-training, reaching up to 38\% in tasks such as tree-cover regression. We attribute this result to the ability of the pre-training to uncover meaningful patterns hidden in the raw data, beneficial for modeling relevant high-level concepts. The pre-trained embeddings emerge as robust representations of regions and trajectories, potentially valuable for a wide range of downstream applications.
\end{abstract}


\section{Introduction}
\label{sec1}

Transformers pre-trained on internet-scale text have significantly advanced the field of Natural Language Processing~(NLP) over the past few years \cite{devlin2018bert, zhang2019ernie, liu2019roberta, lan2019albert, sanh2019distilbert, clark2020electra, he2020deberta, radford2018improving, radford2019language, yang2019xlnet, brown2020language, workshop2022bloom, zhang2022opt, touvron2023llama}. Whether pre-trained to predict the next word in a sentence \cite{radford2018improving, radford2019language, yang2019xlnet, brown2020language, workshop2022bloom, zhang2022opt, touvron2023llama} or unmask randomly masked text \cite{devlin2018bert, zhang2019ernie, liu2019roberta, lan2019albert, sanh2019distilbert, clark2020electra, he2020deberta}, these models learn rich word embeddings that excel at a wide range of language tasks~\cite{devlin2018bert, radford2018improving}.

\begin{figure}
  \centering
  \includegraphics[width=0.9\linewidth]{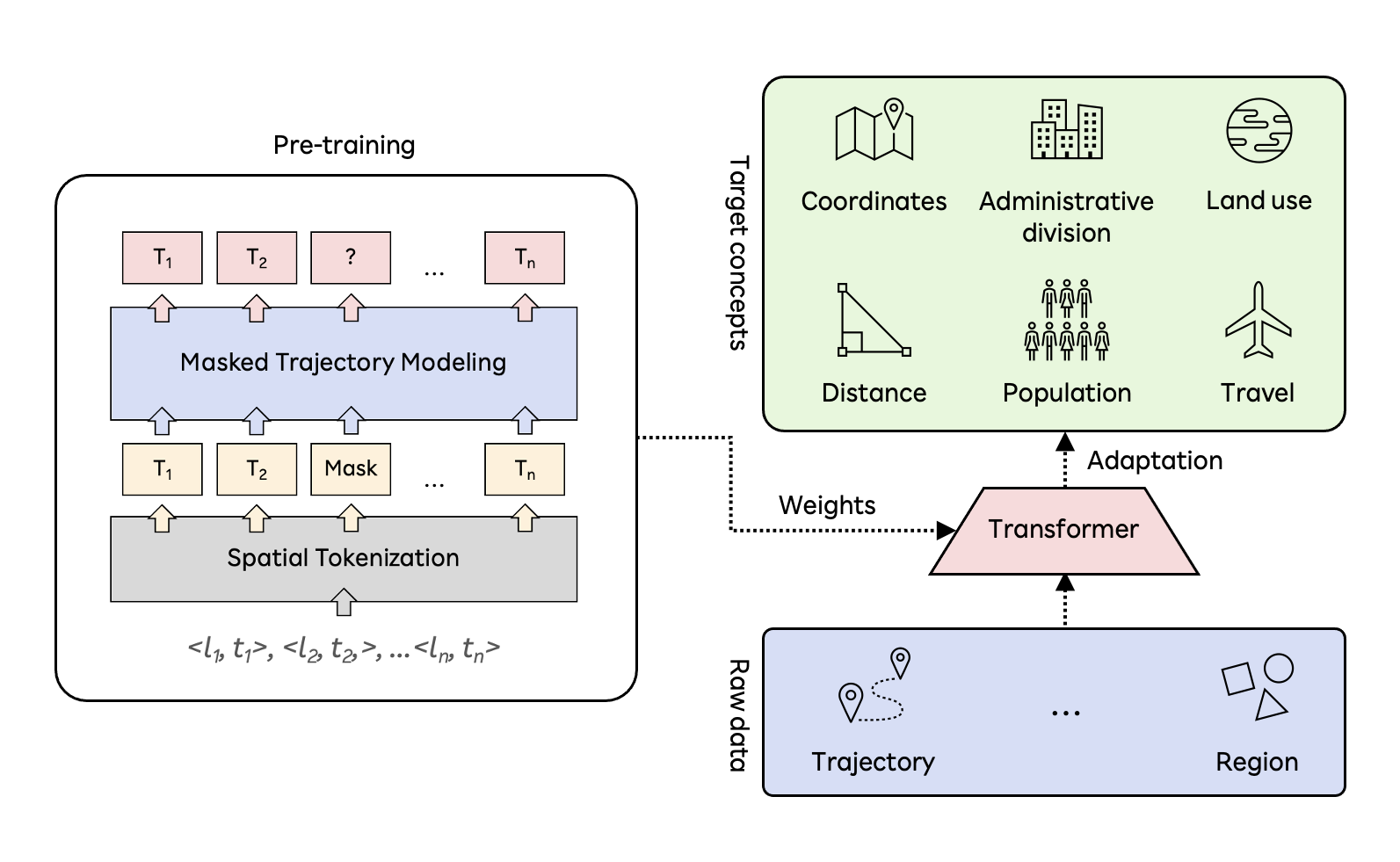}
  \caption{A transformer pre-trained from scratch on country-scale unlabeled human mobility data is adapted to model a variety of high-level concepts manifesting at different levels of spatial analysis.}
  \label{fig1}
\end{figure}

Motivated by their success in NLP and recognizing the inherent parallels between human mobility data and natural language, the geospatial artificial intelligence (GeoAI) community has recently expressed keen interest in pre-trained transformers \cite{musleh2022let, musleh2022towards, musleh2023demonstration, najjar2023towards, xue2022leveraging, xue2023promptcast, zhang2023large, wang2023would, liang2023exploring}. Despite encountering challenges, such as limited pre-training data, these studies have demonstrated the potential of such models for human mobility understanding.

While this recent development marks a significant step forward in the field. It raises questions about whether transformers pre-trained on unlabeled human mobility data \cite{musleh2022let, musleh2022towards, musleh2023demonstration, najjar2023towards} truly capture meaningful mobility patterns, akin to how word embeddings capture structure and semantics in natural language.

In this paper, we attempt to answer this question by thoroughly evaluating embeddings learned by a bidirectional transformer pre-trained via random masking \cite{devlin2018bert} on real-world, country-scale, unlabeled human mobility data to encapsulate a wide array of concepts directly and indirectly related to human mobility. 

Empirical analysis demonstrates that our embeddings not only benefit significantly from pre-training but also, through adaptation, can develop a deep understanding of the target geography and its corresponding mobility patterns.

As a key contribution this paper presents empirical evidence suggesting that self-supervised pre-training of a transformer on country-scale human mobility data uncovers patterns relevant for modeling a wide array of geospatial concepts, such as population count and travel behavior. Which to the best of our knowledge has never been reported before.

The remainder of this paper is organized into four key sections. In Section \ref{sec2}, we provide an overview of previous works that utilize pre-trained transformers for human mobility understanding. Section \ref{sec3} outlines the materials and methods employed in our study, offering insights into data, transformer architecture, pre-training and adaptation approaches used. Section \ref{sec4} presents the evaluation results of our pre-trained embeddings. Finally, in Section \ref{sec5}, we summarize the paper, draw conclusions, and engage in a discussion on our findings and avenues for future research.

\begin{table}
\begin{center}
  \caption{Previous works organized by pre-training data modality.}
  \label{tab1}
  \begin{footnotesize}
  \begin{tabular}{l|l|l}
\toprule
Work                                                  & Modality                                                                         & Task                              \\
\hline
Xue et al \cite{xue2022leveraging, xue2023promptcast} & \multirow{4}{*}{Language}       & POI visitor forecasting           \\ 
Wang et al \cite{wang2023would}                       &                                         & Next-location prediction          \\ 
Zhang et al \cite{zhang2023large}                     &                                         & Anomaly detection                 \\ 
Liang et al \cite{liang2023exploring}                 &                                         & Public event prediction                \\ 
% \small{Xue et al \cite{xue2023promptcast}}                  &                                         & \small{POI visitor forecasting}           \\ 
\hline
Musleh et al \cite{musleh2022let}                     & \multirow{4}{*}{Trajectory}     & \multirow{4}{*}{Task agnostic}    \\ 
Musleh \cite{musleh2022towards}                       &                                         &                                           \\
Musleh \& Mokbel \cite{musleh2023demonstration}       &                                         &                                           \\
Najjar \cite{najjar2023towards}                       &                                         &                                           \\
\bottomrule
\end{tabular}
  \end{footnotesize}
\end{center}
\end{table}

\section{Previous Works}
\label{sec2}

Previous works that utilize \emph{pre-trained} transformers for human mobility understanding can be broadly categorized into two groups: studies utilizing transformers pre-trained on language, i.e., large language models (LLMs) \cite{xue2022leveraging, xue2023promptcast, zhang2023large, wang2023would, liang2023exploring}, and those using transformers pre-trained from scratch on trajectory data \cite{musleh2022let, musleh2022towards, musleh2023demonstration, najjar2023towards}. See Table \ref{tab1} for a summary.

In the first group, human mobility data is translated into natural language prompts which are employed to fine-tune an LLM on a target downstream task. Xue et al were the first to showcase the potential of this approach for human mobility understanding. In their work \cite{xue2022leveraging, xue2023promptcast}, they empirically demonstrated that, for the task of Point-of-interest (POI) visitor forecasting, fine-tuned LLMs perform comparably to specialized models trained exclusively on numerical data. 

Other researchers followed suit, reporting similar results on different human mobility understanding tasks, such as next-location prediction \cite{wang2023would}, trajectory anomaly detection \cite{zhang2023large} and public event prediction \cite{liang2023exploring}.

The second group, on the other hand, is a part of an emergent effort within the GeoAI community aimed at developing Geo Foundation Models \cite{mai2022towards, mai2023opportunities, xue2023artificial, rao2023building, mai2023spatial} for trajectory intelligence. These models are task-agnostic, pre-trained from scratch on extensive trajectory data with the goal of developing a profound understanding of the dynamics of human mobility. 

In the work by Musleh et al \cite{musleh2022let}, a vision for a unified framework for trajectory analysis is proposed. The authors advocate for a model pre-trained once on unlabeled trajectory data, and minimally fine-tuned for a wide range of downstream tasks. This vision is realized in the framework named TrajBERT, as demonstrated by Musleh in~\cite{musleh2022towards}, where a BERT-like transformer, pre-trained on taxi-trip data, learns to impute trajectories. Subsequently, Musleh and Mokbel reported similar results on a larger dataset in \cite{musleh2023demonstration}. Later, Najjar \cite{najjar2023towards} demonstrated that a BERT-like transformer, pre-trained through random masking on user check-in data, not only acquires trajectory unmasking skills but also seems to learn higher order tasks, such as trajectory-user linking \cite{najjar2022trajectory}. 

While our work resembles \cite{musleh2022let, musleh2022towards, musleh2023demonstration, najjar2023towards} in terms of training objective, transformer architecture and data modality, it is to the best of our knowledge the first to present extensive empirical evidence suggesting that a transformer pre-trained on unlabeled human mobility data learns embeddings capable, through fine-tuning, of developing a deep understanding of high-level geospatial attributes, such as population and land cover, of the target study area.

\begin{figure}
  \centering
  \includegraphics[width=0.9\linewidth]{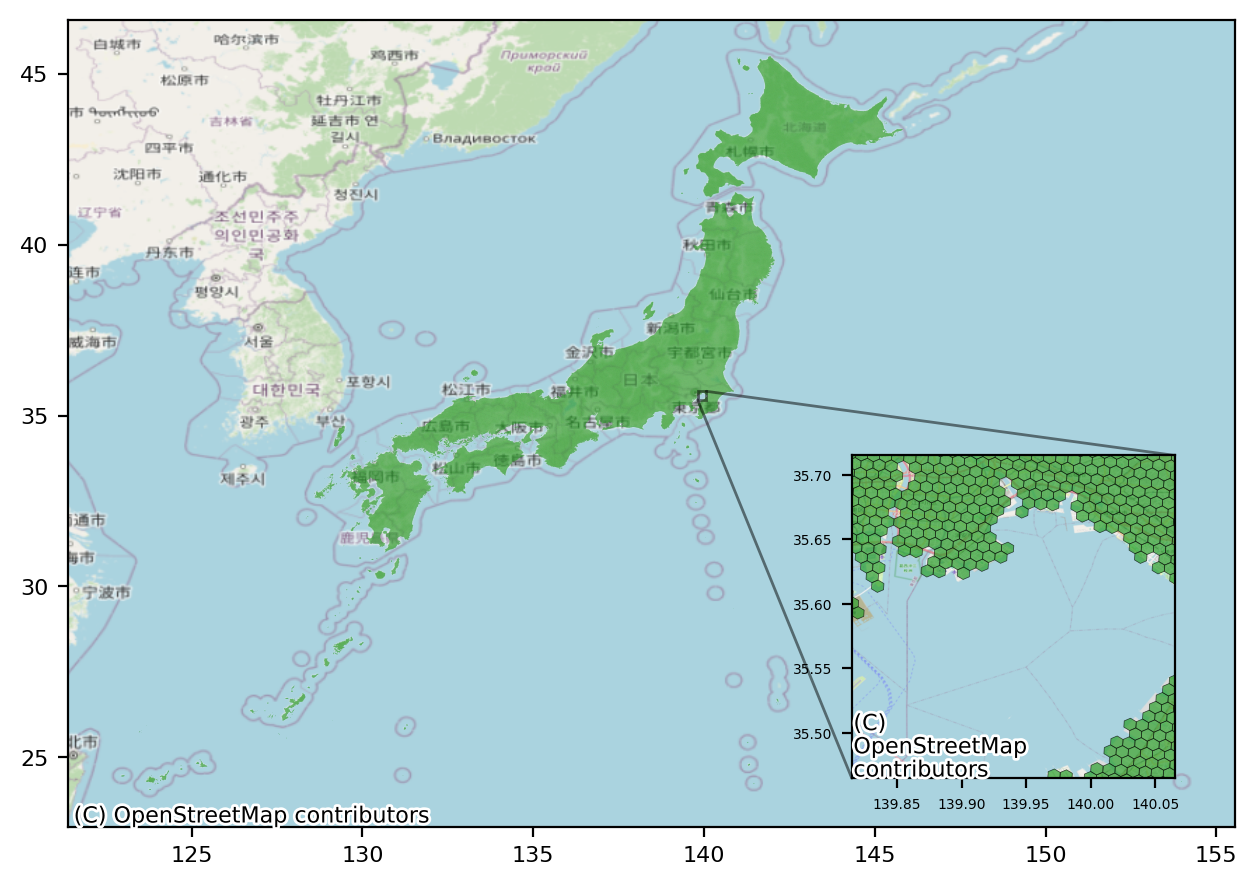}
  \caption{GPS data points aggregated using an Uber H3 grid of resolution 8. Missing polygons indicate data gaps.}
  \label{fig2}
\end{figure}

\section{Materials and Methods}
\label{sec3}

\subsection{Data}

We utilize a proprietary dataset of location data, collected and maintained by Rakuten\footnote{\url{https://www.rakuten.com/}}. The dataset contains over 53 billion GPS data points generated by more than 6 million anonymous users active through out Japan between July 2022 and June 2023. See Figure \ref{fig2} for a visualization of the dataset's spatial coverage. We processed this data into approximately 17 million trajectories, which, to the best of our knowledge, is significantly larger than any other public dataset of its kind \cite{cho2011friendship, yang2016participatory, weeplaces, acmsigcup2017}. We used this dataset to pre-train, validate, and evaluate our model as explained in the following. See Appendix \ref{apxa1} for details on data processing.

\subsection{Transformer}
We adopt BERT \cite{devlin2018bert} as the base architecture for our pre-trained transformer. BERT, which stands for Bidirectional Encoder Representations from Transformers, is a transformer-based model released by Google in 2018. It has revolutionized the field of NLP by introducing a bidirectional approach to language modeling. Unlike unidirectional models, BERT considers context from both directions, enabling it to grasp the meaning of words based on their surrounding context. This bidirectional understanding enhances the model's ability to comprehend nuances, leading to improved performance in various NLP tasks.

Similar to previous works \cite{musleh2022let, musleh2022towards, musleh2023demonstration, najjar2023towards}, we adapt BERT to handle trajectory data by incorporating a spatial-tokenization step which is necessary since transformers are designed to process natural language rather than location data \cite{najjar2023towards}. See Appendix \ref{apxa2} for implementation details.

\subsection{Pre-training}
We leverage Masked Trajectory Modeling (MTM) as a self-supervised pre-training task, a technique previously employed for the same purpose in \cite{najjar2023towards}. A related pre-training task, trajectory imputation, was utilized by Musleh et al in~\cite{musleh2022let, musleh2022towards, musleh2023demonstration}. Both approaches have demonstrated the potential of random masking for learning from unlabeled trajectory data.

Drawing parallels with Masked Language Modeling (MLM) in NLP \cite{devlin2018bert}, MTM involves randomly masking a portion of a trajectory, prompting the model to predict the missing points based on their surrounding context. See Figure \ref{fig1} for an illustration. 

Similar to how MLM enforces an understanding of word contextual relationships in a sentence, MTM enables the model to grasp the contextual relationships between points in a trajectory.

It is noteworthy that, at this stage, our data comprises sequences of hashes (Uber H3) representing user location over a month's time, e.g., \texttt{882e601a5dfffff}, $\cdots$, \texttt{882e601141fffff}.

The process of learning to unmask millions of trajectories presumably forces the model to learn embeddings that capture the contextual relationships between the hashes constituting the target study area. This rationale underscores our choice of employing MTM as a pre-training task. See Appendix \ref{apxa3} for implementation details.

\subsection{Adaptation}
Adaptation is the process of fine-tuning a pre-trained model for a specific task or domain to enhance its performance in that particular area. For example, LLMs are pre-trained on massive amounts of diverse text data to learn general language patterns and semantics. However, fine-tuning allows these models to be customized for a specific application, such as sentiment analysis.

In our context, we employ adaptation to assess the ability of the pre-trained embeddings to encapsulate attributes of the target geography relevant to human mobility, such as population count and land cover. This is achieved by framing the target concept as a downstream task (e.g., population count inference) and comparing the model's performance with and without pre-training. Ideally, pre-training enables the base model to uncover patterns in the raw data that will, in theory, lead to improved downstream performance.

To this end, we utilize three well-known adaptation approaches, namely fine-tuning, few-shot learning and zero-shot learning. See Appendix~\ref{apxa3} for implementation details.

\section{Results}
\label{sec4}

We present the obtained adaptation results organized into two groups based on the model's input whether it is a region or a trajectory. Refer to Appendix \ref{apxa4} for details on evaluation metrics, and pre-training/adaptation implementation.

\subsection{Regions}

Learning embeddings to represent geographic regions is an active area of GeoAI research \cite{spruyt2018loc2vec, du2018zone2vec, jenkins2019unsupervised, wang2020urban2vec, xiang2020region2vec, wozniak2021hex2vec}. In a similar vein, we demonstrate that embeddings pre-trained on randomly masked sequences of (geo) hashes can effectively encode various attributes of the regions these hashes represent. Specifically, we examine the following: geographic location, population count, administrative divisions, and land cover. See Appendix~\ref{apxa3} for details on the datasets used to adapt different models.

\bigbreak \noindent \emph{Geographic location}

\begin{figure}
  \centering
  \includegraphics[width=0.9\linewidth]{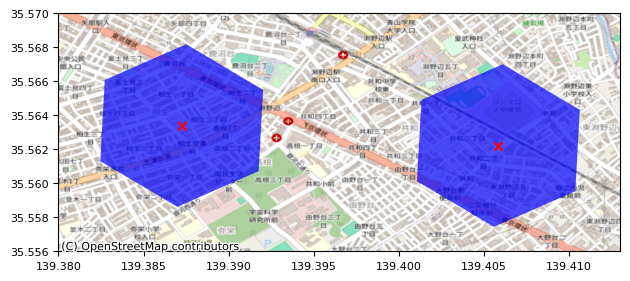}
  \caption{Two example hexagons and their central coordinates marked in red.}
  \label{fig4}
\end{figure}

Geographic location refers to a specific point or area on the Earth's surface identified by its coordinates. Similarly, the geographic location of a given hexagon is determined by the coordinates of its center. In this experiment, we assess the ability of our pre-trained embeddings to encode geographic location. We frame the problem as a multi-target regression with the objective of inferring the central coordinates (latitude/longitude pair) of a given hexagon (see Figure \ref{fig4}).

The obtained results, summarized in Table \ref{tab3}, indicate that pre-training (compared to initialization with random weights) reduces inference error by 0.476 and 0.432 degree in latitude and longitude, respectively. Which translates to 71 km in terms of Haversine distance. This reduction is significant, especially considering that we are using hexagons with an area of 0.74 km$^2$ each. It is worth noting that we used relative coordinates, meaning coordinates were calculated with respect to a reference point, which is the central point of the study area.

\begin{table}
  \caption{Hexagon centroid (Latitude/longitude pair) regression.}
  \label{tab3}
  \begin{center}
  \begin{footnotesize}
      \begin{tabular}{l|cc|cc}
    \toprule
    \multirow{2}{*}{Model}         & \multicolumn{2}{c|}{Latitude}          & \multicolumn{2}{c}{Longitude}          \\
                                   & MAE$\downarrow$  & RMSE$\downarrow$  & MAE$\downarrow$  & RMSE$\downarrow$  \\
    \hline
    Transformer\textsubscript{\tiny{Small}}                          & 0.36              & 0.837              & 0.475             & 1.079              \\
    Transformer\textsubscript{\tiny{Medium}}                          & 0.353             & 0.796              & 0.527             & 1.111              \\
    Transformer\textsubscript{\tiny{Large}}                           & 0.362             & 0.788              & 0.545             & 1.134              \\
    \hline
    BERT\textsubscript{\tiny{Randomly-initialized}}             & 0.564             & 0.886              & 0.53              & 1.109              \\
    BERT\textsubscript{\tiny{Pre-trained}} \tiny{(Zero-shot)}   & 3.015             & 4.006              & 3.679             & 4.715              \\
    BERT\textsubscript{\tiny{Pre-trained}} \tiny{(Few-shots)}   & 2.734             & 3.523              & 3.613             & 4.193              \\
    BERT\textsubscript{\tiny{Pre-trained}} \tiny{(Fine-tuned)}  & \textbf{0.088}    & \textbf{0.123}     & \textbf{0.098}    & \textbf{0.159}     \\
    \bottomrule
    \end{tabular}
\end{footnotesize}
\end{center}
\end{table}

We attribute this result to the pre-training objective, wherein learning to unmask sequences of hashes enabled the model to capture the hierarchical and spatial relationships between the hexagons that constitute the study area thereby enhancing its ability to infer their geographic locations relative to one another.

\bigbreak \noindent \emph{Population count}

Population count refers to the total number of individuals within a specific geographic area. In the following, we assess our model's ability to infer population at the hexagon level. To this end, we utilize WorldPop \cite{worldpop2018} data aggregated at a 1 km resolution as a ground truth. Since this data is not directly measured but rather collected using remote sensing methods, we frame the target task as a 10-class (deciles) classification rather than direct population count regression.

\begin{figure}
  \centering
  \includegraphics[width=0.9\linewidth]{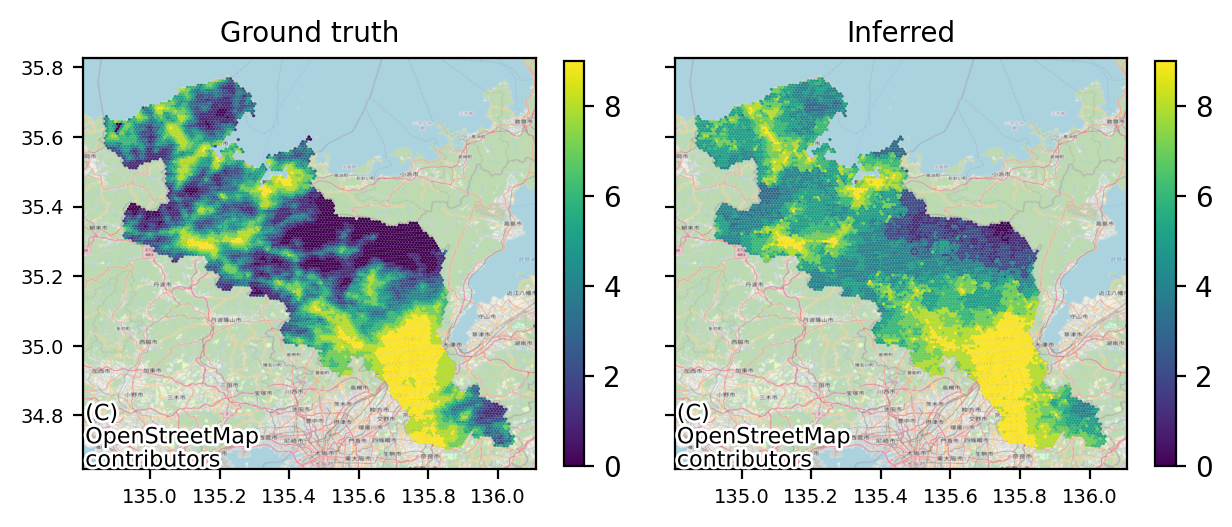}
  \caption{Population (deciles) map of Kyoto prefecture: Ground-truth and inferred.}
  \label{fig5}
\end{figure}

While the obtained results (refer to Table \ref{tab4}) do not suggest any of the models perform well at the task at hand, they clearly indicate that pre-training is beneficial, with overall gain of 0.115 in F-Score~(33\% performance gain). Furthermore, a map of the inferred population (see Figure \ref{fig5}) illustrates the model's ability to preserve~(to a certain degree) the spatial distribution of population over the study area.

\begin{figure}
  \centering
  \includegraphics[width=0.9\linewidth]{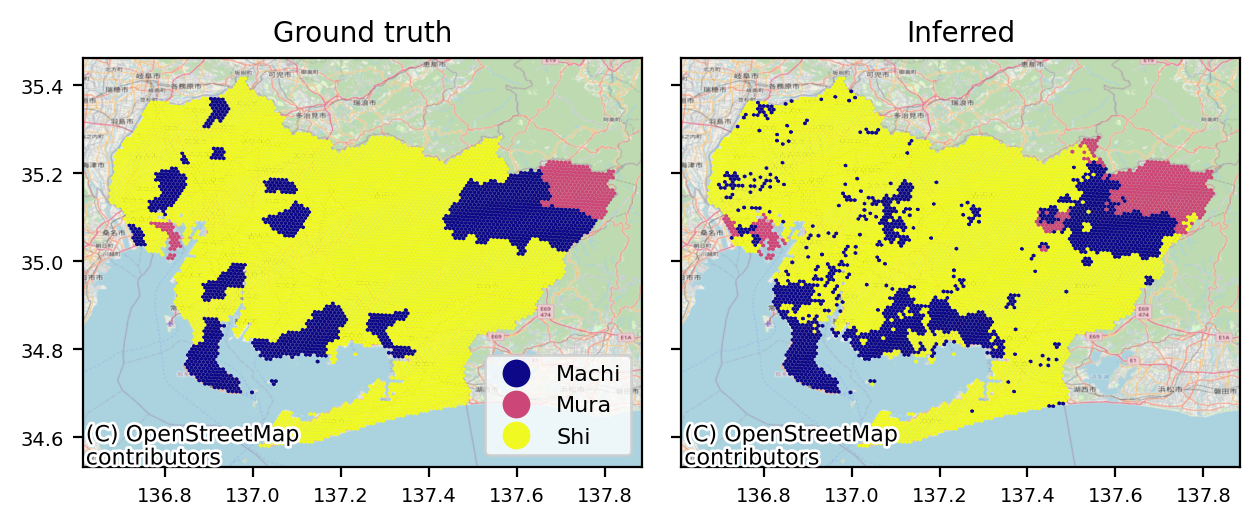}
  \caption{Municipality classification map of Aichi prefecture: Ground-truth and inferred.}
  \label{fig7}
\end{figure}

Given the intrinsic link between human mobility and population distribution, we attribute this result to the pre-training's ability to uncover patterns underlying in the raw data indicative of population. We hypothesize that this capability emerges as an indirect result of learning to unmask trajectories at a large scale.

\bigbreak \noindent \emph{Administrative divisions}

Administrative divisions refer to the hierarchical structure by which a geographic area is organized into smaller units for the purpose of governance and public administration. Administrative divisions are decided based on a complex interplay of factors, including geography, population, culture, politics, and history. 

Japan, for example, consists of eight main regions divided into~47 prefectures. Municipalities within each prefecture are assigned one of three broad categories: Shi (city), Machi (town), or Mura (village) \cite{wiki_admin_div_japan}. 

% \begin{figure}
%     \centering
%     \begin{subfigure}{}
%     \centering
%         \centering
%         \includegraphics[height=1.25in]{figures/hexagon-prefecture.png}
%         % \caption{}
%     \end{subfigure}
%     \begin{subfigure}{}
%     \centering
%         \centering
%         \includegraphics[height=1.25in]{{figures/hexagon-region.png}}
%         % \caption{}
%     \end{subfigure}
%     \caption{Japan's hexagons labeled by their corresponding prefecture (Up) and regions (Bottom).}
%   \label{fig6}
% \end{figure}

Through a series of experiments, we evaluate how well our pre-trained embeddings can encapsulate attributes such as prefecture, region, and municipality. In all three cases, the problem is framed as hexagon classification with the objective of assigning a given hexagon its prefecture, region and municipality classification, respectively. We used GADM dataset \cite{gadm2024} as a ground truth.

Results are summarized in Tables \ref{tab5}, \ref{tab6}, and \ref{tab7}. The impact of pre-training is clearly noticeable in all three scenarios, with an overall performance boost ranging between 5\% and 12\% across the three classification tasks. Refer to Figure \ref{fig7} for a subjective evaluation of the best model performance at municipality classification.

\begin{table}
\begin{center}
  \caption{Hexagon-level prefecture classification.}
  \label{tab5}
  \begin{footnotesize}
  \begin{tabular}{l|c|c|c|c}
\toprule
Model                          & Acc.$\uparrow$  & P$\uparrow$  & R$\uparrow$  & F$\uparrow$ \\
% \midrule
\hline
Transformer\textsubscript{Small}                          & 0.8                  & 0.866                 & 0.862              & 0.863              \\
Transformer\textsubscript{Medium}                          & 0.79                 & 0.858                 & 0.857              & 0.856              \\
Transformer\textsubscript{Large}                          & 0.791                & 0.861                 & 0.859              & 0.857              \\
% \midrule
\hline
BERT\textsubscript{Randomly-initialized}                      & 0.79                 & 0.86                  & 0.857              & 0.857              \\
% \hline
BERT\textsubscript{Pre-trained} (Zero-shot)     & 0.024                & 0.003                 & 0.017              & 0.003              \\
BERT\textsubscript{Pre-trained} (Few-shots)                               & 0.021                & 0.034                 & 0.185              & 0.058              \\
BERT\textsubscript{Pre-trained} (Fine-tuned)                               & \textbf{0.958}       & \textbf{0.969}        & \textbf{0.968}     & \textbf{0.968}     \\
\bottomrule
\end{tabular}
\end{footnotesize}
\end{center}
\end{table}

\begin{table}
\begin{center}
  \caption{Hexagon-level region classification.}
  \label{tab6}
  \begin{footnotesize}
  \begin{tabular}{l|c|c|c|c}
\toprule
Model                          & Acc.$\uparrow$  & P$\uparrow$  & R$\uparrow$  & F$\uparrow$ \\
% \midrule
\hline
Transformer\textsubscript{Small}                           & 0.932                & 0.939                 & 0.938              & 0.938              \\
Transformer\textsubscript{Medium}                          & 0.933                & 0.94                  & 0.937              & 0.938              \\
Transformer\textsubscript{Large}                          & 0.934                & 0.942                 & 0.936              & 0.938              \\
% \midrule
\hline
BERT\textsubscript{Randomly-initialized}                      & 0.934                & 0.942                 & 0.94               & 0.941              \\
% \hline
BERT\textsubscript{Pre-trained} (Zero-shot)     & 0.123                & 0.03                  & 0.129              & 0.037              \\
BERT\textsubscript{Pre-trained} (Few-shots)                               & 0.125                & 0.046                 & 0.178              & 0.054              \\
BERT\textsubscript{Pre-trained} (Fine-tuned)                               & \textbf{0.994}       & \textbf{0.994}        & \textbf{0.994}     & \textbf{0.994}     \\
\bottomrule
\end{tabular}
\end{footnotesize}
\end{center}
\end{table}

\begin{table}
\begin{center}
  \caption{Hexagon-level population classification.}
  \label{tab4}
\begin{footnotesize}
  \begin{tabular}{l|c|c|c|c}
\toprule
Model                             & Acc.$\uparrow$  & P$\uparrow$  & R$\uparrow$  & F$\uparrow$ \\
% \midrule
\hline
Transformer\textsubscript{Small}                   & 0.358                & 0.361                 & 0.361              & 0.358              \\
Transformer\textsubscript{Medium}                   & 0.358                & 0.354                 & 0.362              & 0.354              \\
Transformer\textsubscript{Large}                   & 0.359                & 0.359                 & 0.361              & 0.357              \\
% \midrule
\hline
BERT\textsubscript{Randomly-initialized}         & 0.353                & 0.342                 & 0.356              & 0.344              \\
% \hline
BERT\textsubscript{Pre-trained} (Zero-shot)      & 0.11                 & 0.04                  & 0.113              & 0.041              \\
BERT\textsubscript{Pre-trained} (Few-shots)      & 0.143                & 0.16                  & 0.144              & 0.101              \\
BERT\textsubscript{Pre-trained} (Fine-tuned)     & \textbf{0.46}        & \textbf{0.46}         & \textbf{0.462}     & \textbf{0.459}     \\
\bottomrule
\end{tabular}
\end{footnotesize}
\end{center}
\end{table}

\begin{table}
\begin{center}
  \caption{Hexagon-level municipality classification.}
  \label{tab7}
  \begin{footnotesize}
  \begin{tabular}{l|c|c|c|c}
\toprule
Model                          & Acc.$\uparrow$  & P$\uparrow$  & R$\uparrow$  & F$\uparrow$ \\
% \midrule
\hline
Transformer\textsubscript{Small}                          & 0.782                & 0.78                  & 0.778              & 0.775              \\
Transformer\textsubscript{Medium}                          & 0.784                & 0.783                 & 0.781              & 0.779              \\
Transformer\textsubscript{Large}                          & 0.786                & 0.784                 & 0.783              & 0.781              \\
% \midrule
\hline
BERT\textsubscript{Randomly-initialized}                      & 0.783                & 0.783                 & 0.779              & 0.776        \\
% \hline
BERT\textsubscript{Pre-trained} (Zero-shot)     & 0.333                & 0.116                 & 0.34               & 0.173              \\
BERT\textsubscript{Pre-trained} (Few-shots)                               & 0.333                & 0.118                 & 0.343              & 0.175              \\
BERT\textsubscript{Pre-trained} (Fine-tuned)                               & \textbf{0.849}       & \textbf{0.846}        & \textbf{0.846}     & \textbf{0.844}     \\
\bottomrule
\end{tabular}
\end{footnotesize}
\end{center}
\end{table}

Given the inherent role human mobility (people's movement) plays in shaping administrative divisions, we attribute this gain to the model's ability to capture the spatial patterns and population flows that underpin the delineation and organization of administrative units--a proficiency refined through its pre-training to unmask millions of trajectories.

\bigbreak \noindent \emph{Land cover}

Land cover refers to the various natural and artificial features present on the Earth's surface, including vegetation, water bodies, and human-made structures.

We evaluate the ability of our model to infer the distribution of two land cover classes, namely built-up areas and tree cover. We utilize a global land-cover land-use dataset sampled at a 10 m resolution \cite{karra2021}. We labeled each hexagon by its ratio of the target land cover and framed the problem as regression. The results are summarized in Tables \ref{tab8} and \ref{tab9}.

\begin{table}
\begin{center}
  \caption{Hexagon-level built-up area regression.}
  \label{tab8}
  \begin{footnotesize}
  \begin{tabular}{l|c|c|c}
\toprule
Model                          & MAE$\downarrow$     & RMSE$\downarrow$  & R$^2$$\uparrow$   \\
% \midrule
\hline
Transformer\textsubscript{Small}                          & 0.106                & 0.183              & 0.535              \\
Transformer\textsubscript{Medium}                          & 0.103                & 0.183              & 0.535              \\
Transformer\textsubscript{Large}                          & 0.102                & 0.182              & 0.538              \\
% \midrule
\hline
BERT\textsubscript{Randomly-initialized}                      & 0.107                & 0.18               & 0.549              \\
% \hline
BERT\textsubscript{Pre-trained} (Zero-shot)     & 0.173                & 0.317              & -0.399             \\
BERT\textsubscript{Pre-trained} (Few-shots)                               & 0.178                & 0.269              & -0.007             \\
BERT\textsubscript{Pre-trained} (Fine-tuned)                               & \textbf{0.089}       & \textbf{0.148}     & \textbf{0.696}     \\
\bottomrule
\end{tabular}
  \end{footnotesize}
\end{center}
\end{table}

\begin{figure}
  \centering
  \includegraphics[width=0.9\linewidth]{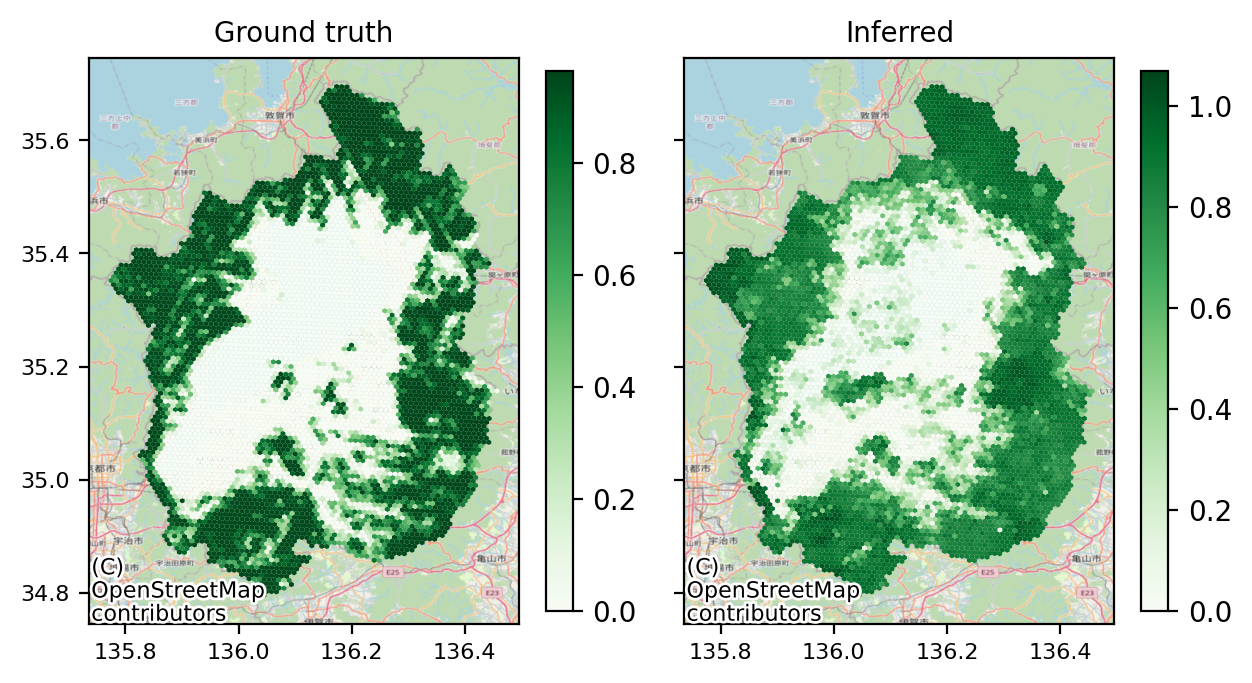}
  \caption{Tree-cover map of Shiga prefecture: Ground-truth and inferred.}
  \label{fig8}
\end{figure}

\begin{figure}
  \centering
  \includegraphics[width=0.9\linewidth]{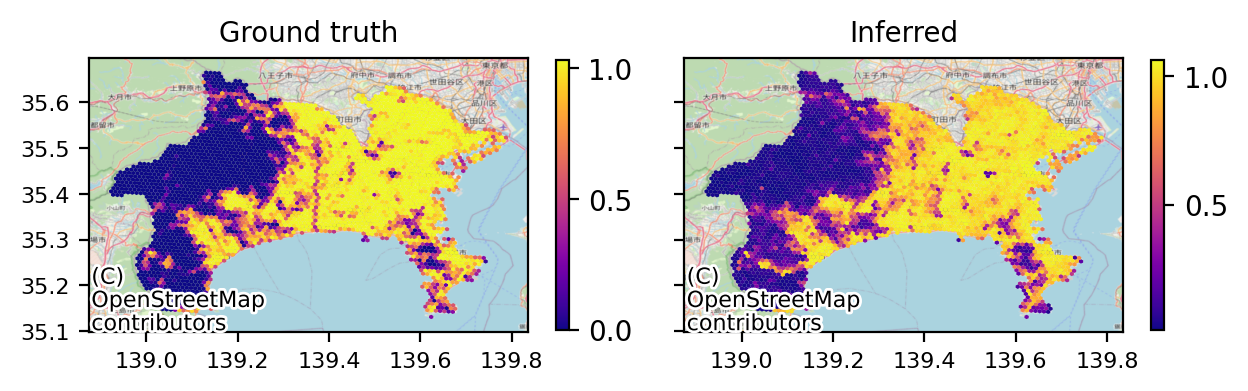}
  \caption{Built-up area map of Kanagawa prefecture: Ground-truth and inferred.}
  \label{fig9}
\end{figure}

In both cases, pre-training significantly boosts performance, with an overall gain of 0.147 to 0.158 in R$^2$ score which translates to 26\% to 38\% improvement in variance explanation. For a subjective evaluation we mapped in Figures \ref{fig8} and \ref{fig9} the inferred distribution of the respective land cover classes along side ground-truth data. The model's ability to preserve the spatial distribution is clearly illustrated in both examples.

\begin{table}
\begin{center}
  \caption{Hexagon-level tree cover regression.}
  \label{tab9}
  \begin{footnotesize}
  \begin{tabular}{l|c|c|c}
\toprule
Model                          & MAE$\downarrow$     & RMSE$\downarrow$  & R$^2$$\uparrow$   \\
% \midrule
\hline
Transformer\textsubscript{Small}                           & 0.211                & 0.282              & 0.431              \\
Transformer\textsubscript{Medium}                           & 0.204                & 0.282              & 0.433              \\
Transformer\textsubscript{Large}                           & 0.202                & 0.283              & 0.427              \\
% \midrule
\hline
BERT\textsubscript{Randomly-initialized}                      & 0.201                & 0.286              & 0.415              \\
% \hline
BERT\textsubscript{Pre-trained} (Zero-shot)     & 0.58                 & 0.652              & -2.041             \\
BERT\textsubscript{Pre-trained} (Few-shots)                               & 0.293                & 0.373              & 0.005              \\
BERT\textsubscript{Pre-trained} (Fine-tuned)                               & \textbf{0.164}       & \textbf{0.244}     & \textbf{0.573}     \\
\bottomrule
\end{tabular}
  \end{footnotesize}
\end{center}
\end{table}

Here again, we attribute the model's performance gain to pre-training by random masking, which we hypothesize is capable of uncovering sequential and spatial patterns underlying the raw data indicative of these two land cover classes. For example, heavily forested areas tend to have different human mobility patterns than areas dense with man-made structures.

\emph{It is worth mentioning that when tested with land cover classes such as water bodies, our pre-trained embeddings not only performed poorly, but also did not benefit from pre-training. A result that highlights the limitation of our embeddings to encapsulate geospatial attributes weakly correlated with human mobility.}

\begin{figure}
  \centering
  \includegraphics[width=0.9\linewidth]{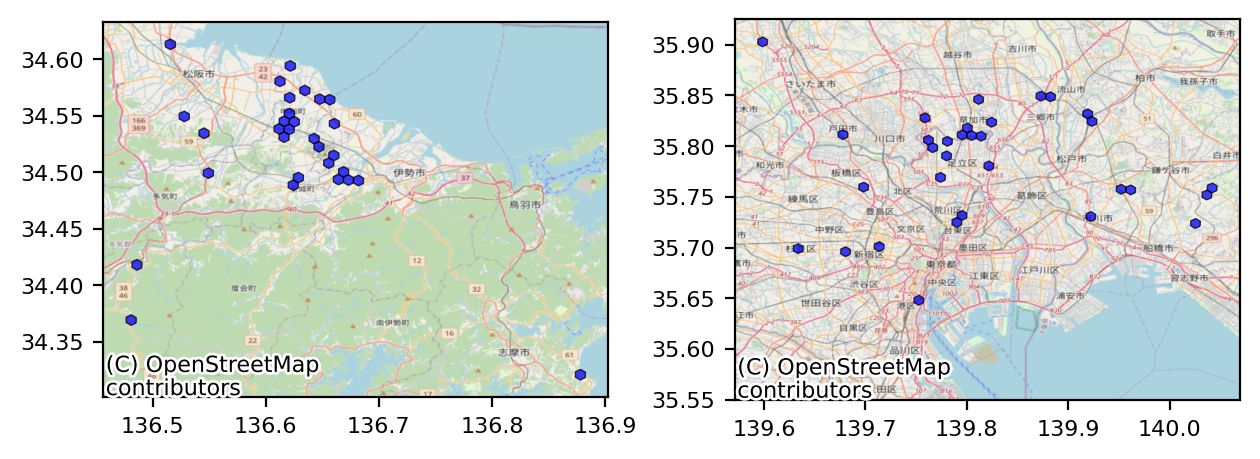}
  \caption{Two example trajectories with varying geographic diversity scores: Trajectory on the left has a score of 1 while the one on the right has a score of 3.}
  \label{fig10}
\end{figure}

\subsection{Trajectories}

In this section, we empirically demonstrate that our pre-trained embeddings can effectively encapsulate trajectory attributes, such as geographic diversity, and trip length. Moreover, we extend the analysis to examine user attributes manefisting at the trajectory level, such as travel behavior, and exposure to greenery.

Unlike the previous experiments where the model's input is a region (hexagon), in the following we work with trajectories, i.e., sequences of hexagons.

\begin{table}
\begin{center}
  \caption{Geographic diversity regression.}
  \label{tab10}
  \begin{footnotesize}
  \begin{tabular}{l|c|c|c|c}
\toprule
Model                          & MAE$\downarrow$  & MAPE$\downarrow$  & RMSE$\downarrow$  & R$^2$$\uparrow$ \\
% \midrule
\hline
Transformer\textsubscript{\tiny{Small}}                          & 0.703             & 0.448              & 0.99               & 0.296            \\
Transformer\textsubscript{\tiny{Medium}}                          & 0.675             & 0.398              & 0.991              & 0.295            \\
Transformer\textsubscript{\tiny{Large}}                          & 0.65              & 0.354              & 0.997              & 0.285            \\
% \midrule
\hline
BERT\textsubscript{Randomly-initialized}                      & 0.705             & 0.427              & 1.023              & 0.249            \\
% \hline
BERT\textsubscript{\tiny{Pre-trained}} \tiny{(Zero-shot)}     & 1.979             & 1.119              & 2.305              & -2.817           \\
BERT\textsubscript{\tiny{Pre-trained}} \tiny{(Few-shots)}                             & 0.711             & 0.452              & 1.017              & 0.258            \\
BERT\textsubscript{\tiny{Pre-trained}} \tiny{(Fine-tuned)}                               & \textbf{0.082}    & \textbf{0.039}     & \textbf{0.225}     & \textbf{0.964}   \\
\bottomrule
\end{tabular}
\end{footnotesize}
\end{center}
\end{table}

\begin{table}
\begin{center}
  \caption{Average trip length regression.}
  \label{tab11}
  \begin{footnotesize}
  \begin{tabular}{l|c|c|c|c}
\toprule
Model                          & MAE$\downarrow$  & MAPE$\downarrow$  & RMSE$\downarrow$  & R$^2$$\uparrow$ \\
% \midrule
\hline
Transformer\textsubscript{\tiny{Small}}                          & 9.191             & 2.57               & 23.239             & 0.2              \\
Transformer\textsubscript{\tiny{Medium}}                          & 10.826            & 4.417              & 22.865             & 0.226            \\
Transformer\textsubscript{\tiny{Large}}                          & 9.487             & 2.864              & 23.143             & 0.207            \\
% \midrule
\hline
BERT\textsubscript{Randomly-initialized}                      & 10.675            & 3.824              & 23.916             & 0.153              \\
% \hline
BERT\textsubscript{\tiny{Pre-trained}} \tiny{(Zero-shot)}       & 11.39             & 0.992              & 28.359             & -0.191           \\
BERT\textsubscript{\tiny{Pre-trained}} \tiny{(Few-shots)}                              & 9.749             & 2.365              & 26.839             & -0.067           \\
BERT\textsubscript{\tiny{Pre-trained}} \tiny{(Fine-tuned)}                              & \textbf{2.147}    & \textbf{0.642}     & \textbf{9.769}     & \textbf{0.859}   \\
\bottomrule
\end{tabular}
\end{footnotesize}
\end{center}
\end{table}

\begin{table*}
\begin{center}
  \caption{Cross-prefecture and cross-region travel (binary) classification.}
  \label{tab12}
  \begin{footnotesize}
  \begin{tabular}{l|cccc|cccc}
\toprule
\multirow{2}{*}{Model}         & \multicolumn{4}{c|}{Cross-prefecture}                             & \multicolumn{4}{c}{Cross-region}  \\
                               & Accuracy $\uparrow$  & Precision $\uparrow$  & Recall $\uparrow$  & F-Score $\uparrow$  & Accuracy $\uparrow$  & Precision $\uparrow$  & Recall $\uparrow$  & F-Score $\uparrow$     \\
% \midrule
\hline
Transformer\textsubscript{Small}                            & 0.694                & 0.7                   & 0.698              & 0.696               & 0.65                 & 0.803                 & 0.823              & 0.805              \\
Transformer\textsubscript{Medium}                            & 0.702                & 0.703                 & 0.703              & 0.703               & 0.634                & 0.804                 & 0.825              & 0.8                \\
Transformer\textsubscript{Large}                            & 0.709                & 0.711                 & 0.711              & 0.71                & 0.603                & 0.795                 & 0.819              & 0.784              \\
% \midrule
\hline
BERT\textsubscript{Randomly-initialized}                     & 0.704                & 0.718                 & 0.71               & 0.705               & 0.644                & 0.809                 & 0.829              & 0.806        \\
% \hline
BERT\textsubscript{Pre-trained} (Zero-shot)    & 0.5                  & 0.227                 & 0.477              & 0.308               & 0.483                & 0.648                 & 0.289              & 0.272              \\
BERT\textsubscript{Pre-trained} (Few-shots)                               & 0.653                & 0.656                 & 0.656              & 0.654               & 0.5                  & 0.838                 & 0.797              & 0.707                  \\
BERT\textsubscript{Pre-trained} (Fine-tuned)                               & \textbf{0.971}       & \textbf{0.972}        & \textbf{0.972}     & \textbf{0.971}      & \textbf{0.989}       & \textbf{0.993}        & \textbf{0.993}     & \textbf{0.993}            \\
\bottomrule
\end{tabular}
\end{footnotesize}
\end{center}
\end{table*}

\begin{table}
\begin{center}
  \caption{User exposure to greenery regression.}
  \label{tab13}
  \begin{footnotesize}
  \begin{tabular}{l|c|c|c}
\toprule
Model                          & MAE$\downarrow$   & RMSE$\downarrow$  & R$^2$$\uparrow$   \\
% \midrule
\hline
Transformer\textsubscript{Small}                          & 0.044              & 0.084              & 0.531              \\
Transformer\textsubscript{Medium}                          & 0.045              & 0.082              & 0.553              \\
Transformer\textsubscript{Large}                          & 0.043              & 0.083              & 0.542              \\
% \midrule
\hline
BERT\textsubscript{Randomly-initialized}                      & 0.043              & 0.083              & 0.541              \\
% \hline
BERT\textsubscript{Pre-trained} (Zero-shot)     & 0.129              & 0.163              & -0.755             \\
BERT\textsubscript{Pre-trained} (Few-shots)                               & 0.07               & 0.12               & 0.048              \\
BERT\textsubscript{Pre-trained} (Fine-tuned)                               & \textbf{0.028}     & \textbf{0.06}      & \textbf{0.762}     \\
\bottomrule
\end{tabular}
\end{footnotesize}
\end{center}
\end{table}

\bigbreak \noindent \emph{Geographic diversity}

Geographic diversity of a trajectory refers to the variety of locations traversed through out the trajectory, potentially reflecting lifestyle, habits, or general mobility patterns. We quantify geographic diversity by calculating the number of unique prefectures spanned per trajectory (see Figure \ref{fig10}). We frame the problem as a regression task, with the objective of inferring the number of unique prefectures given a trajectory. 

The obtained results (refer to Table \ref{tab10}) show significant performance improvement from pre-training, with an R$^2$ score boost of~0.715. This task builds upon prefecture classification, so the performance gain from pre-training aligns with expectations. However, the extent of this improvement is intriguing. We attribute it to BERT's ability to leverage context, which is lacking when working with regions, e.g., prefecture classification.

\bigbreak \noindent \emph{Trip length} 

We define trip length as the average distance between two successive points given a trajectory. To calculate trip length, we average the Haversine distance between the centers of successive hexagons, measured in kilometers. Changes in trip length can reflect shifts in lifestyle, such as commute patterns or physical activity levels.

In this experiment, we evaluate the effectiveness of our pre-trained embeddings in encapsulating trip length. Our findings, summarized in Table \ref{tab11}, demonstrate a notable gain in R$^2$ of over 0.7~(or 8.5 km reduction in inference error) as a result of pre-training. We attribute this performance boost to the pre-training's ability to capture spatial relationships among the hexagons constituting the target study area, as evidenced in the geographic location experiment presented earlier. \bigbreak

Moving on to trajectory-derived user attributes, in the following we focus on travel behavior, and exposure to greenery.

\bigbreak \noindent \emph{Travel behavior}

We are interested in two travel patterns: cross-prefecture and cross-region travel. Travel, like other offline activities, has been the focus of previous studies \cite{najjar2023you}. To ensure that we exclude commuting, we define travel as crossing prefecture/region borders during weekends and/or national holidays. To this end, we frame the problem as a simple binary classification task, aiming to infer whether the user is a traveler based on their activity over a month.

The obtained results, summarized in Table~\ref{tab12}, not only indicate that our embeddings effectively encapsulate both travel patterns but also demonstrate that pre-training significantly enhances the model’s performance by up to 0.26 in F-Score. This finding adds to the evidence that random masking applied to trajectory data is highly effective in learning the administrative divisions of the target study area.

\begin{figure}
  \centering
  \includegraphics[width=0.9\linewidth]{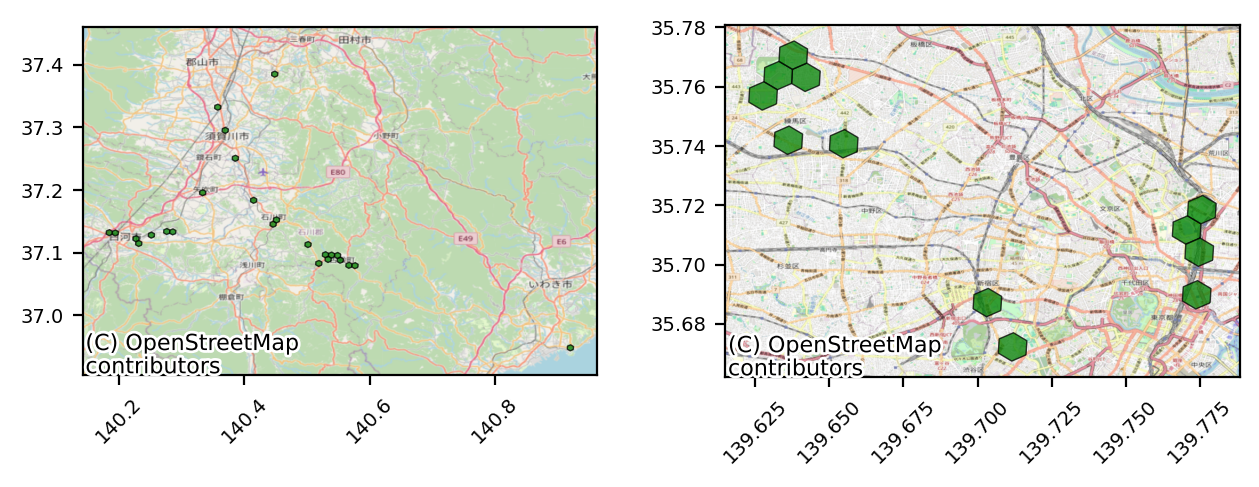}
  \caption{Trajectories of two different users and their exposure to greenery scores: Trajectory on the left has a score of~0.63 while the one on the right has a score of 0.04.}
  \label{fig11}
\end{figure}

\bigbreak \noindent \emph{Exposure to greenery}

We define exposure to greenery as the user's proximity to tree-covered areas over time. We quantify it by averaging the tree-cover\footnote{Here we used the same land cover dataset used in the previous section.} ratios of the hexagons visited by the user over a month (see Figure~\ref{fig11}). This score may offer insights into an individual's potential access to nature and its associated benefits for well-being and environmental engagement.

Next, we evaluate our model's capacity to infer the user's exposure to greenery, as depicted in Table \ref{tab13}. Results indicate that pre-training has improved the model's performance by over 40\% in terms of R$^2$ score.

It is important to acknowledge the limitations of our definition of greenery exposure. Greenery encompasses more than just tree cover; grassy lands, pastures, bushes, etc., are integral parts of green spaces. Furthermore, proximity to tree-covered areas doesn't ensure actual \quotes{exposure} to greenery; individuals may remain indoors for the duration.

\bigbreak 

We argue that the results of the previous two experiments only scratch the surface on the potential of pre-trained embeddings as user representations for a variety of applications, including offline activity recommendation, and location-based user clustering.

\section{Discussion}
\label{sec5}

\bigbreak \noindent \emph{Overview}

In this paper, we presented empirical evidence suggesting that a transformer pre-trained on millions of unlabeled human-mobility trajectories learns embeddings capable, through fine-tuning, of developing a deep understanding of the target geography and its corresponding mobility patterns. Utilizing an adaptation framework, we evaluated the performance of our pre-trained embeddings in encapsulating a wide array of geospatial concepts directly and indirectly related to human mobility. The obtained results indicated a significant performance boost gained from pre-training. We attribute this result to the ability of the pre-training to uncover meaningful patterns hidden in the raw human mobility data, beneficial for modeling relevant high-level concepts.

We argue that the findings we report in this paper not only advance our understanding of the capabilities of pre-trained transformers for human mobility understanding but also opens avenues for improved applications in domains such as urban planning, transportation, and socio-economic analysis.

\bigbreak \noindent \emph{Results}

We adopted BERT as the base architecture for our pre-trained transformer. Our transformer underwent pre-training via masked trajectory modeling applied to 17 million anonymized trajectories generated by over 6 million users active throughout Japan over a period of 12 months. After 40 epochs of pre-training validation perplexity dropped from few thousand to less than 4. To evaluate the efficacy of the pre-trained embeddings, we adapted our model to encapsulate region and trajectory-related attributes, such as population count and geographic diversity. Furthermore, we extended the analysis to infer user attributes, including travel behavior, and exposure to greenery. Empirical analysis showed consistent performance boost gained from pre-training, demonstrating the potential of the pre-trained embeddings as generic region and/or trajectory representations. 

It is worth noting that our experiments also demonstrate the importance of fine-tuning in leveraging the full potential of pre-training. In fact, pre-trained models adapted through zero-shot or few-shot learning consistently underperformed a fine-tuned model initialized with random weights.

\bigbreak \noindent \emph{Why it works}

We attribute the efficacy of pre-training via masked trajectory modeling to the following factors:

\begin{enumerate}
\item Context understanding: Transformers are adept at capturing contextual relationships in sequences. By pre-training a transformer on millions of randomly masked trajectories, the model is forced to develop an understand of the spatial context of the hashes that make up the trajectories. Which translates to learning embeddings that encode the spatial relationships of the hexagons that constitute the target study area.
\item Statistical correlations: Human mobility is inherently linked to high-level concepts such as population and land cover. For instance, densely populated areas tend to have different mobility patterns compared to sparsely populated ones. Similarly, various types of land cover influence human movement and mobility patterns.
\end{enumerate}

In summary, although it may initially appear counterintuitive, it is evident that a transformer pre-trained on country-scale human mobility data can learn embeddings capable of effectively encapsulating high-level concepts, such as population, land cover and administrative divisions, that describe the target study area.

\bigbreak \noindent \emph{Privacy implications}

Location data is inherently personal, encapsulating our habits, preferences, and vulnerabilities. When ethically harnessed, Artificial Intelligence applied to location data presents significant opportunities across various domains, including location-based recommendations, urban planning, and emergency response systems. However, the ubiquity of location data, coupled with accessible algorithmic advancements, makes it vulnerable to misuse by untrustworthy parties.

In our study, we utilized user location data collected at a large scale, and demonstrated enhanced performance across multiple applications, including travel behavior classification and inference of exposure to certain land cover classes. 

To protect user privacy, we aggregated location data using a mid-resolution spatial grid. Additionally, we systematically eliminated user-identifying features such as names, precise addresses, and income levels from our analysis.

We acknowledge the potential for misuse by untrustworthy parties, such as inferring users' mental health based on their exposure to greenery, with a model like ours. Therefore, despite the promising opportunities AI presents when applied to location data, we underscore the critical importance of responsible stewardship and handling of user location data. It is imperative to ensure that privacy and ethical considerations remain paramount in both its utilization and development.

\bigbreak \noindent \emph{Future work}

While our approach has demonstrated success, avenues for further advancements remain open. Integrating time into pre-training represents a primary avenue for future work. Currently, our model is spatially aware but lacks explicit consideration of the temporal dimension. This could be achieved simply by utilizing temporal embeddings, learned end to end, fused with the spatial embeddings at the input level. Incorporating time will enable a deeper understanding of the dynamics of human mobility. This enhancement promises to elevate predictive capabilities and provide insights into the changes in mobility across time. 

Integrating various privacy-preserving elements into the model development life cycle is another critical area for future research. While we have ensured the anonymization of the used data, our aim is to investigate the integration of techniques such as differential privacy at the pre-training stage. 

In conclusion, our research highlights two critical avenues for future advancements: the integration of temporal embeddings into pre-training to better understand human mobility dynamics and the incorporation of privacy-preserving techniques into the model development process. These endeavors aim to improve predictive capabilities while maintaining privacy standards within the GeoAI research community.

\section*{Acknowledgment}
The author would like to thank Kyle Mede for his valuable feedback on the final version of the manuscript.

\bibliographystyle{IEEEtran}
\bibliography{bibs}

\appendix
\section{Appendix}
\label{apxa}

\subsection{Data processing}
\label{apxa1}

\subsubsection{From GPS coordinates to trajectories} Initially, we projected the GPS data points onto an Uber H3 grid \cite{brodsky2018h3} of resolution 8 (Spatial grid of hexagon-shaped cells of an approximate area of 0.74 km$^2$ each). Subsequently, sequences of data points are transformed into check-ins, defined as clusters of points belonging to the same user occurring within a 10-minute period and confined to the same cell. We constructed trajectories for each user by chronologically concatenating their cell hashes generated over a month's time. To enhance data quality, we excluded trajectories with fewer than 10 check-ins or those with less than 3 unique grid cells. Additionally, for privacy considerations, we excluded any user identifying information, such user ID, from the final data.

\subsubsection{Data splits} We randomly split the (17 million) trajectories into three subsets: 70\% for pre-training, 15\% for validation and 15\% for testing. All the downstream (adaptation) evaluations presented in this manuscript are based on subsets of the test split, which remains as a whole unseen by the model during pre-training.

\subsection{Transformers}
\label{apxa2}

\subsubsection{BERT} We adopted the Hugging Face \cite{wolf2019huggingface} implementation of BERT as the base transformer\footnote{See Table \ref{tab14} for details on hyper-parameters.}. Additionally, we experimented with two variations of BERT: 1) BERT initialized with random weights (BERT\textsubscript{Randomly-initialized}), and 2) BERT pre-trained on trajectory data (BERT\textsubscript{Pre-trained}) in three different flavors: zero-shot, few-shots and fine-tuned. 

All BERT variants were adapted using the exact same datasets. The only difference between BERT\textsubscript{Randomly-initialized} and BERT\textsubscript{Pre-trained} is that the former is initialized with random weights while the latter is pre-trained from scratch on country-scale trajectory data.

\subsubsection{Baselines} We used the Hugging Face library to implement all three baselines (Transformer\textsubscript{Small, Medium \& Large}) detailed in Table \ref{tab2}. \emph{We limited our baselines to transformer based since transformers represent state-of-the-art in sequential data modeling}. In all experiments we trained baselines in a fully supervised fashion \emph{using the exact same datasets utilized to fine-tune the BERT-based model}.

\begin{table}
\begin{center}
  \caption{Baseline configuration.}
  \label{tab2}
  \begin{footnotesize}
  \begin{tabular}{lcccc}
\toprule
Baseline   & layer \#  & head \#  & Input dim.    & Hidden dim.     \\
\hline
Transformer\textsubscript{Small} & 3             & 12           & 768           & 3072            \\
Transformer\textsubscript{Medium} & 6             & 12           & 768           & 3072            \\
Transformer\textsubscript{Large} & 9             & 12           & 768           & 3072            \\
\bottomrule
\end{tabular}
\end{footnotesize}
\end{center}
\end{table}

\subsubsection{Spatial tokenization} We utilized the the spatial tokenizer implemented in \cite{najjar2023towards} using Uber H3 grid of resolution 8 as an encoder and WordPiece \cite{wu2016google} for a tokenizer trained from scratch with a vocabulary of 30522 tokens.

\subsection{Pre-training and adaptation}
\label{apxa3}

\subsubsection{Pre-training} First, whole-hash pre-masking with a 20\% masking ratio is applied to both the training and validation data splits. Next, BERT was pre-trained, to minimize perplexity \cite{rabiner1989tutorial}, for 40 epochs (Roughly 240 hours). Pre-training was done in parallel on two NVIDIA Tesla V100 GPU chips with 128G of memory each. See Table \ref{tab14} for details on hyper-parameters.

\subsubsection{Adaptation} To adapt a transformer-based model (BERT variants or baselines) we added on top a classification or regression layer (Depending on the task) initialized with random weights and subsequently trained the model on hand-labeled randomly-selected adaptation dataset for 20 epochs. The adaptation dataset consists of 50k samples\footnote{Depending on the target downstream task, a sample is either a hexagon randomly sampled from within Japan's level-zero administrative boundaries or a trajectory randomly sampled from the test split.} randomly split into 80\% training and 20\% testing subsets. The whole training split is used for fine-tuning, while a random 64 samples are used for few-shot learning. No training is conducted in zero-shot learning. The same machine used for pre-training was used for adaptation. See Table \ref{tab14} for details on hyper-parameters.

It is worth mentioning that the dataset used to fine-tune BERT is exactly the same one used to train the previously mentioned transformer baselines.

\begin{table}
\begin{center}
  \caption{Hyper-parameters configuration.}
  \label{tab14}
  \begin{footnotesize}
  \begin{tabular}{l|ccc}
\toprule
\multirow{2}{*}{Hyper-parameter}         & \multirow{2}{*}{Pre-training}  &  \multicolumn{2}{c}{Adaptation}    \\
                        &                                &  Trajectory  & Region              \\
\hline
Training epochs         & 40                             & 20           & 20             \\
Learning rate           & 5e-5          & 2e-5                    & 2e-5           \\
Weight decay            & 1e-1          & 1e-1                    & 1e-1           \\
Per-device batch size   & 64            & 32                      & 1024             \\
Optimizer               & AdamW         & AdamW                   & AdamW          \\
\hline
Sequence max length     & 512           & 512                     & 512      \\
Truncation              & \checkmark    & \checkmark              & -    \\
Padding                 & \checkmark    & \checkmark              & -     \\
\hline
Random seed             & 123           & 123                     & 123            \\
\bottomrule
\end{tabular}
\end{footnotesize}
\end{center}
\end{table}

\subsection{Evaluation}
\label{apxa4}

\subsubsection{Evaluation metrics} To evaluate the quality of pre-training, we used perplexity \cite{rabiner1989tutorial} as a metric. Perplexity measures how "surprised" the model is when it sees new data and is a commonly used metric in NLP in general and masked language modeling specifically. 

To evaluate the quality of adaptation, we used Accuracy, Precision, Recall, and F-Score\footnote{To account for label imbalance we used weighted-average Accuracy, Precision, Recall and F-Score metrics.} for classification tasks, and Mean Absolute Error (MAE), Mean Absolute Percentage Error (MAPE), Root Mean Squared Error (RMSE), and R$^2$ for regression tasks. These are well-known evaluation metrics widely used for both classification and regression, respectively.

\subsubsection{Pre-training} 
\begin{figure}
  \centering
  \includegraphics[width=0.9\linewidth]{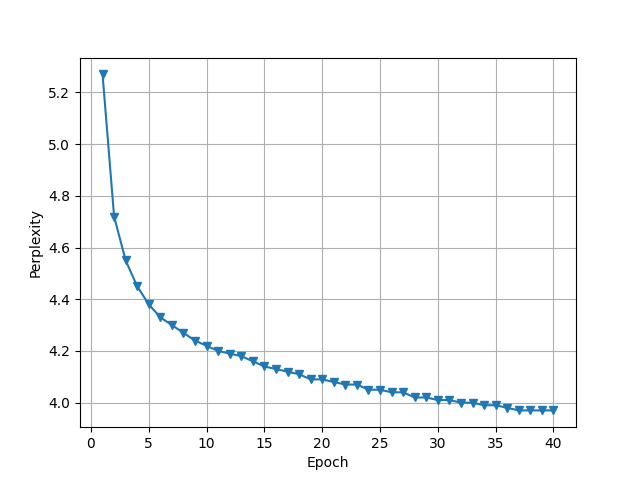}
  \caption{Pre-training evaluation: Validation perplexity plotted over time.}
  \label{figx}
\end{figure}

After 40 epochs of pre-training, validation perplexity reaches 3.79, which is a significant drop from the initial (random initialization) value of 2327 (See Figure \ref{figx}). It is worth noting that by the end of the first epoch, the model has already seen over a billion spatial tokens (hash or sub-hash), which might explain the relatively low perplexity value of 5.27. We further investigated this theory by pre-training a different model using a significantly smaller dataset of roughly 80 million tokens, and we found that the perplexity gradually drops to a comparable value~($\approx8$) by the end of the 12th epoch. This supports our hypothesis regarding the initially relatively low perplexity value.

\subsubsection{Adaptation} 
\begin{figure}
  \centering
  \includegraphics[width=0.9\linewidth]{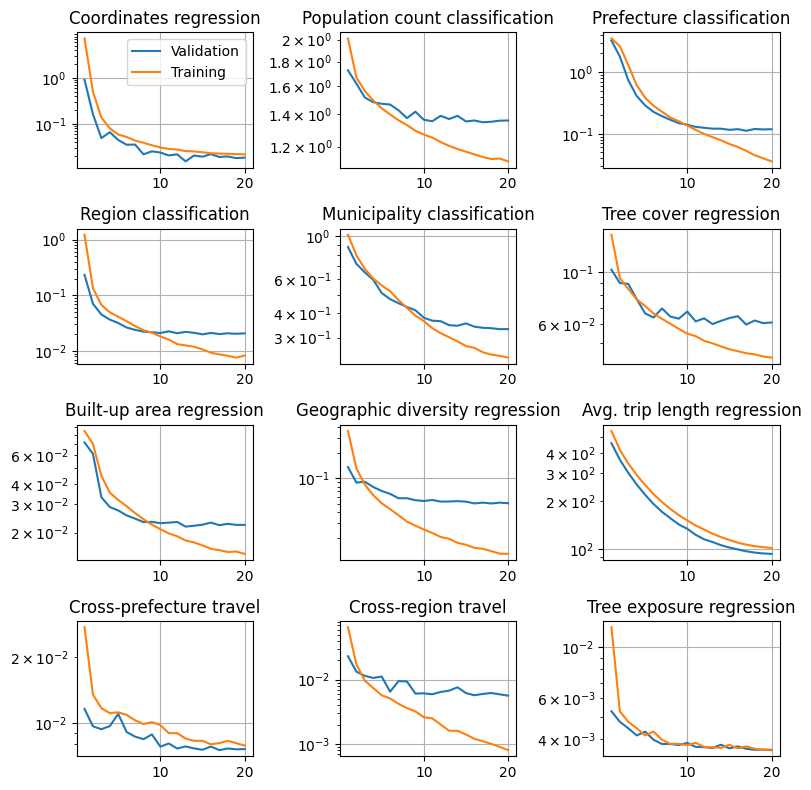}
  \caption{Adaptation evaluation: Validation and training loss curves (fine-tuning). X and Y axes represent epoch and loss, respectively.}
  \label{figy}
\end{figure}

We reported the best results (Before overfitting) over 20 epochs while closely watching the loss curves for any signs of overfitting. Fine-tuning loss curves of all downstream tasks are shown in Figure \ref{figy}.

\end{document}